\documentclass[preprint,12pt]{elsarticle}

\usepackage{amssymb}
\usepackage{lineno}
\usepackage[english]{babel}
\usepackage{listings}
\usepackage[autostyle,english=american]{csquotes}
\MakeOuterQuote{"}
\usepackage[acronym,toc]{glossaries}
\usepackage{makecell}
\usepackage{url}
\newacronym{MOOSE}{\texttt{MOOSE}}{Multiphysics Object Oriented Simulation Environment}
\newacronym{pde}{PDE}{partial differential equation}
\newacronym{fem}{FEM}{finite element method}
\newacronym{mpi}{MPI}{Message Passing Interface}
\newacronym{petsc}{PETSc}{Portable, Extensible Toolkit for Scientific Computation}
\newacronym{ad}{AD}{automatic differentiation}
\newacronym{gui}{GUI}{graphical user interface}
\newacronym{jfnk}{JFNK}{Jacobian-free Newton-Krylov}
\newacronym{sqa}{SQA}{Software Quality Assurance}

\journal{SoftwareX}

\graphicspath{
{figures/}
}

\usepackage{hyphenat}
\usepackage{xspace}
\newcommand{\sys}[1]{\texttt{#1}\xspace}

\begin{document}
\begin{frontmatter}



\title{MOOSE: Enabling Massively Parallel Multiphysics Simulation}

\cortext[cor1]{Corresponding author}
\author[inlmail]{Cody J. Permann\corref{cor1}}
\author[inlmail]{Derek R. Gaston\corref{cor1}}
\author[inlmail]{David Andr\v{s}}
\author[inlmail]{Robert W. Carlsen}
\author[inlmail]{Fande Kong}
\author[inlmail]{Alexander D. Lindsay}
\author[inlmail]{Jason M. Miller}
\author[akselos]{John W. Peterson}
\author[inlmail]{Andrew E. Slaughter}
\author[icesmail]{Roy H. Stogner}
\author[inlmail]{Richard C. Martineau}

%\author[inlmail]{Daniel Schwen}

\address[inlmail]{Computational Frameworks, Idaho National Laboratory, Idaho Falls, ID, 83415}
\address[icesmail]{Institute for Computational and Engineering Sciences, The University of Texas at Austin, Austin, TX, 78712}
\address[akselos]{Akselos Inc., 2101 CityWest Blvd., Houston, TX, 77042}

\begin{abstract}
Harnessing modern parallel computing resources to achieve complex multiphysics simulations is a daunting task.  The \gls{MOOSE} aims to enable such development by providing simplified interfaces for specification of partial differential equations, boundary conditions, material properties, and all aspects of a simulation without the need to consider the parallel, adaptive, nonlinear, finite-element solve that is handled internally.  Through the use of interfaces and inheritance, each portion of a simulation becomes reusable and composable in a manner that allows disparate research groups to share code and create an ecosystem of growing capability that lowers the barrier for the creation of multiphysics simulation codes.  Included within the framework is a unique capability for building multiscale, multiphysics simulations through simultaneous execution of multiple sub-applications with data transfers between the scales.  Other capabilities include automatic differentiation, scaling to a large number of processors, hybrid parallelism, and mesh adaptivity.  To date, \gls{MOOSE}-based applications have been created in areas of science and engineering such as nuclear physics, geothermal science, magneto-hydrodynamics, seismic events, compressible and incompressible fluid flow, microstructure evolution, and advanced manufacturing processes.
\end{abstract}

\end{frontmatter}

%\linenumbers


% Description of your software in maximum 5 pages for first Original Software Publication –- see suggested format;

\section{Motivation and significance}
\label{sec:intro}
%Enabling scientists to build state-of-the-art, scalable finite element simulation tools drives development of the \acrfull{MOOSE}, a framework providing common interfaces to application developers saving time, ensuring correctness and providing advanced simulation capabilities.

As computing capability increases, scientists are exploring computational solutions to complex problems involving multiple scientific domains.  These multiphysics problems tie together many disciplines and involve collaborators with specific knowledge.  Traditionally, multiphysics simulation was achieved by connecting individual simulation tools built by various researchers, which is time-consuming, error-prone, and problematic for taking advantage of modern parallel computing. The \acrfull{MOOSE} \cite{gaston2009moose} offers a better approach for development of such software.  The framework provides a plug-in infrastructure that simplifies definitions of physics, material properties, and postprocessing.  The design allows developers to focus on their scientific endeavor without needing to understand the intricacies of modern parallel computing.  In addition, uniformity in problem specification leads to maximizing reuse. \gls{MOOSE}-based projects can share capabilities without requiring additional development.  This ideal is achieved by a design pattern where community-developed physics "modules" and all \gls{MOOSE}-based applications are libraries and can be linked, enabling conglomerate applications to be created from existing capability.  This concept opens new possibilities for collaborating across development teams to build powerful simulation tools that can be seamlessly combined. This strategy has been a paradigm shift in multiple scientific disciplines\cite{gaston2014physics}.

\section{Software description}
\label{sec:framework}

\subsection{Software architecture}
\label{sec:arch}
%Give a short overview of the overall software architecture.
\gls{MOOSE} is designed to facilitate the creation of production \gls{fem} tools for running high-fidelity, multiphysics simulations. The software is composed of "systems," each one providing an extension point for defining simulation characteristics.  These systems communicate using interfaces, decoupling them to allow for greater code reuse. Each system in \gls{MOOSE} has a specific C++ base class from which application developers inherit from and extend to perform a desired calculation overriding virtual methods to perform unique calculations needed for that application. This approach has two advantages; the foremost is that the application developers have complete control over their calculation using standard C++, should a need to deviate from the prescribed template arise. Secondly, the inheritance chain can be extended beyond the baseline provided by the framework to provide rich capabilities for downstream applications.  \gls{MOOSE} has approximately 40 systems that may be extended. Individually discussing each of these is beyond the scope of this paper, but the systems that provide the core functionality of the framework can be separated into three groups: \gls{pde} related terms, material properties, and in-situ postprocessing.

Two systems for defining \gls{pde} terms are the \sys{Kernel} and \sys{Boundary\-Condition} systems. \sys{Kernel} objects define volumetric integral terms while \sys{BoundaryCondition} objects define surface integral terms that arise from the derivation of the \gls{fem} weak form\cite{fish2007first}. For example, the C++ code snippet in Listing \ref{src:kernel} is an example of an implementation of an advection term for a \gls{pde}. This example illustrates the \sys{Kernel} system, which supports \gls{ad} (see Section \ref{sec:function}). A single method must be extended, \sys{precomputeQpResidual}.

\begin{lstlisting}[caption=Code snippet showing \sys{Kernel} inheritance for the computation of an advective term within a \gls{pde}., frame=tlrb, basicstyle=\tiny, label=src:kernel]
template <ComputeStage compute_stage>
ADReal
Advection<compute_stage>::precomputeQpResidual()
{
  return _velocity[_qp] * _grad_u[_qp];
}
\end{lstlisting}

This is a powerful approach to defining equation terms.  By defining one method, this advection operator can be used in 1D, 2D, or 3D.  It can utilize a constant velocity, couple to a field computed through solving another \gls{pde}, utilize a field computed from another application, or even use a velocity field from an experiment.  This \sys{Kernel} will work on one processor or 100,000 without modification.  Any application needing this term can reuse this object without needing to recode it.  Finally, it can also be tested in a controlled calculation to ensure correctness.

The \sys{Material} system allows for the definition of "material properties," often representing coefficients in a \gls{pde}. These properties may be nonlinear, and may themselves depend on variables in the \gls{pde}. This dependence can be propagated automatically through to the system Jacobian matrix using the \gls{ad} capability of the framework. Once defined, material properties may be consumed by other systems in the framework through a straight-forward producer/consumer model. \gls{MOOSE} then ensures those properties are computed when needed by those objects. Material properties can couple to and depend on other material properties; inter-dependencies are tracked, and calculations are executed in the correct order. This design decouples physics calculations from the coefficients they require, allowing for properties to be shared and reused within and across applications without the need to edit code objects consuming the property.

The term "in-situ postprocessing," a perceived oxymoron, describes a set of systems used for data computation that is typically performed post-simulation which are computed along with other calculations after, during or in-between solves. Systems in this category include the \sys{Postprocessor}, \sys{VectorPostprocessor}, and \sys{AuxKernel} systems. The \sys{Postprocessor} and \sys{VectorPostprocessor} systems compute scalar and vector values, respectively (e.g., total heat flux through a side or the total chemical concentration). These calculations can themselves depend on other variables, material properties, or other postprocessed values. The \sys{AuxKernel} system allows the calculation of "field" or spatially varying data using finite element basis functions. In-situ postprocessing enables calculations to be performed in parallel while running at scale.  These values can be fed back into other systems, including the calculation of \gls{pde} terms or material properties.

The system and interface architecture of the framework allows application developers to build simulation tools capable of solving anything from basic single physics problems to extensive, tightly coupled, cross-disciplinary multiphysics problems only achievable with multiple development teams. The resulting applications, regardless of complexity, are easily extensible and useful for research teams and analysts alike.

\subsection{Software functionalities}
\label{sec:function}
\subsubsection{Parallelism}
\begin{figure}[!htbp]
    \centering
    \includegraphics[width=0.5\textwidth]{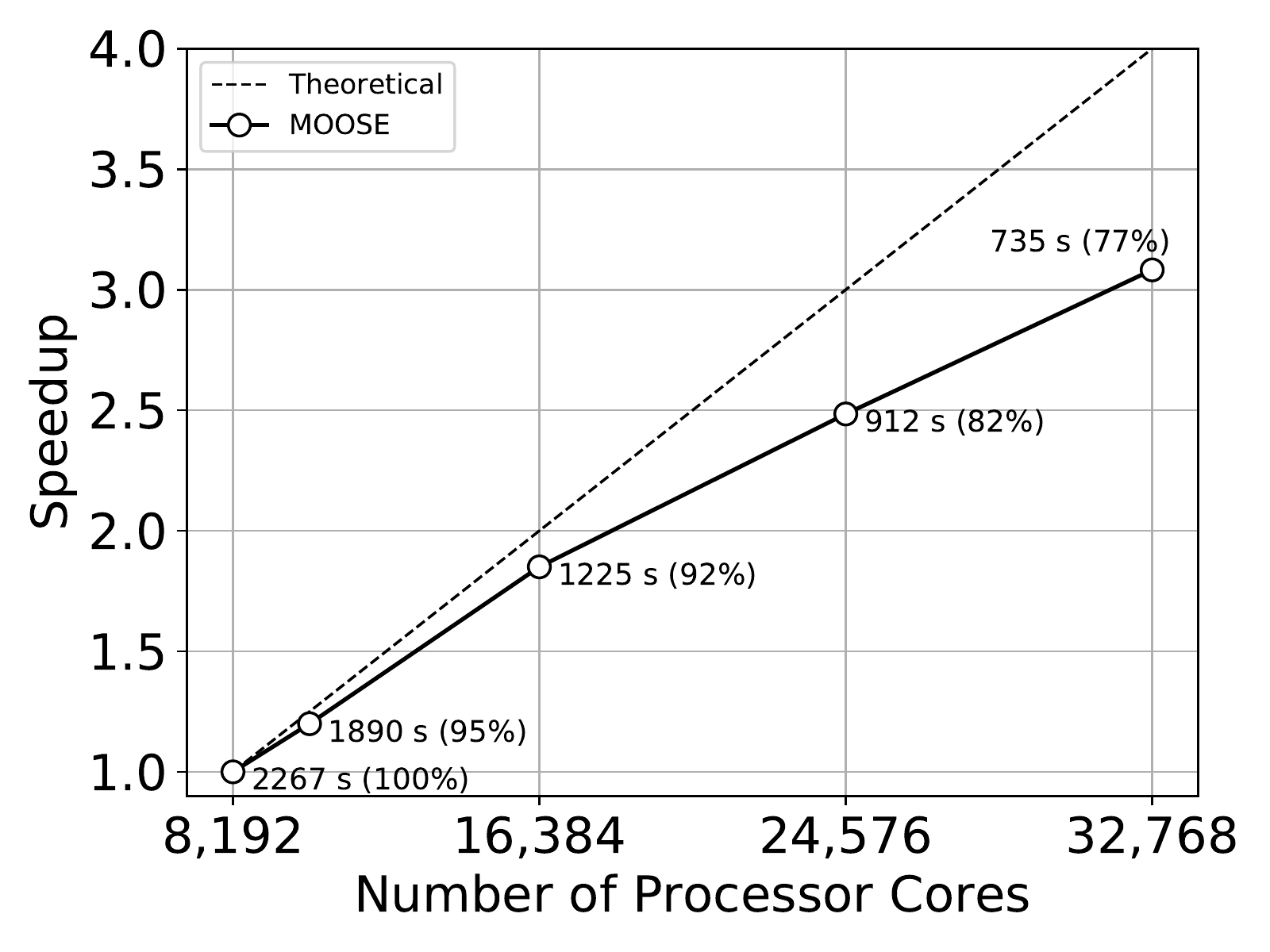}
    \caption{Scaling study of \gls{MOOSE} using up to 32,768 processor cores. Each data point is annotated with the total compute time and the parallel efficiency.  The total computed time does not include the input/ouptut cost and the mesh preparation.}
    \label{fig:scaling}
\end{figure}

\gls{MOOSE} is designed to support highly parallel calculations involving many physical phenomena. The parallel capability is implemented using a hybrid approach consisting of \gls{mpi} calls with optional shared-memory threading available on-node. Traditionally, parallel computing is complicated, involving many low-level serialization, deserialization, and explicit communication routines. \gls{MOOSE} provides a comprehensive abstraction to parallelism, requiring minimal knowledge (if any) of parallel constructs from application developers. Additionally, the framework supports an advanced mesh "pre-splitting" capability that automatically takes into account necessary "ghosting" needed by contact, mortar, patch recovery, or user-defined, non-local requirements. Pre-split meshes can then be read efficiently in parallel, reducing startup time and memory use for very large simulations. The pre-split capability has been used to keep per-core memory consumption reasonable for billion element meshes\cite{gastondissertation}. Scalability of the framework has been demonstrated to over 30,000 processor cores on modern supercomputers for multiple applications including microstructure evolution using the phase-field method and neutron transport. For example, a scaling study is shown in Fig.~\ref{fig:scaling} for a 3D grain growth model based on phase-field equations. The phase-field equations are discretized with 25 grains and 9 order parameters using a first-order Lagrange finite element method on a 3D mesh with 78,643,200 hexahedral elements and 79,300,033 nodes. The resulting system has 713,700,297 unknowns and is solved by the \gls{jfnk} \cite{knoll2004jacobian} method, employing GMRES \cite{saad1986gmres} and a Schwarz preconditioner \cite{kong2016highly, smith2004domain}. The test is conducted using processor counts between 8,192 and 32,768 on the Theta supercomputer at Argonne National Laboratory \cite{anl-theta}. Further details of this test can be found in \cite{kong2018general, kong2018fully}.

\subsubsection{Automatic Differentiation (AD)}
Efficient solution of nonlinear systems of equations using Newton's method requires accurate computation of the Jacobian matrix or its action on a vector \cite{knoll2004jacobian}. When this matrix is approximated via matrix-free methods, accurate preconditioning may still be required for efficient linear convergence. Generating an accurate Jacobian matrix or preconditioner can be a difficult and error-prone task. To reduce the burden on application developers, \gls{MOOSE} has implemented forward-mode \gls{ad}. Using this capability, application developers only implement residual statements, reducing the amount of code and time needed to implement a new physics object. Sharing physics code among applications can consequently be done with greater confidence. In this way, \gls{ad} enhances the modularity of \gls{MOOSE}-based applications.

\subsubsection{Sub-applications}
The architecture of \gls{MOOSE} allows separately developed physics simulation tools to be integrated seamlessly via the \sys{MultiApp} system.  As an example, two different research groups might each be focused on different aspects of a problem: one on an engineering scale simulation and one on microstructure evolution.  Using \gls{MOOSE}, both applications can be developed in a uniform, yet flexible way.  The two research groups may work independently, focusing on their respective scales.  When a higher fidelity, coupled simulation is needed, these two applications can be compiled together, and a multiscale simulation can be performed using the unique \sys{MultiApp} and \sys{Transfer} systems without the need to develop additional code to link the applications together (see \cite{gaston2014physics, martineau2019multiphysics}).

\subsubsection{Restart}
The framework includes a restart system that allows for all aspects of a simulation to be stored in memory or output to checkpoint files. This allows for single applications to be recovered after an unexpected failure (e.g., a power outage), to be restored while conducting iterative solves across multiple applications (see Sub-applications above), as well as to create complex multi-stage initialization for large problems. Notably, this system is extendable so that user-defined data structures can be stored and reloaded when needed in each of these scenarios and when running at scale.

\subsubsection{Testing, Visualization, and Documentation}
\label{sec:python}
A modular framework alone is not sufficient to create robust and useful scientific simulation tools. \gls{MOOSE} also includes infrastructure for testing, visualization, and documentation, which, collectively with the framework, is referred to as the "\gls{MOOSE} Platform." As with the framework, each of these utilities is designed with similar architectures and to be modular and extendable, allowing application developers to customize these tools. The testing system includes tools for creating unit, regression, and system tests, including tests for error conditions and parallel consistency. The \gls{gui} is designed to work with derivative applications containing extended objects and custom syntax. The visualization core of the \gls{gui} may be scripted to create complex visualizations. Finally, the documentation system operates with single-source markdown files to create websites, presentations, or \LaTeX{} documents and is capable of checking cross-reference consistency as documentation or source code changes.

% I DON'T THINK WE NEED THIS SECTION
%\section{Implementation and Empirical Results}
%\label{sec:implementation}
%Implementation details.
%Empirical results.
%Conduct empirical studies and provide results.
%Compare with state-of-the-art software if any, kindly cite relevant work.

\section{Illustrative Example}
\label{sec:example}
%Optional: you may include one explanatory video that will appear next to your article, in the right hand side panel. (Please upload any video as a single supplementary file with your article. Only one MP4 formatted, with 50MB maximum size, video is possible per article. Recommended video dimensions are 640 $\times$ 480 at a maximum of 30 frames/second. Prior to submission please test and validate your .mp4 file at $ http://elsevier-apps.sciverse.com/GadgetVideoPodcastPlayerWeb/verification$. This tool will display your video exactly in the same way as it will appear on ScienceDirect.).

\begin{figure}[!htbp]
    \centering
    \includegraphics[width=\textwidth]{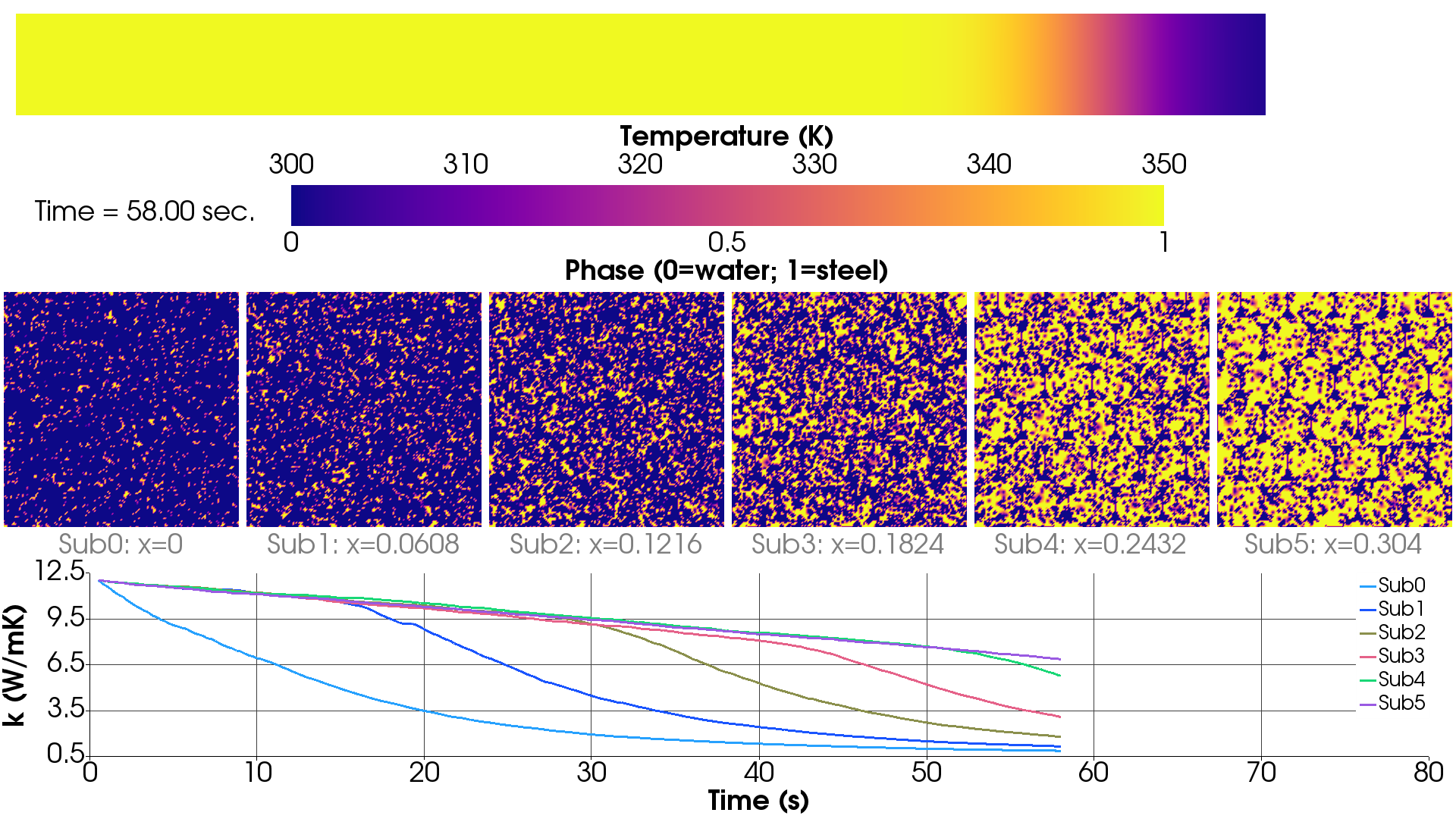}
    \caption{Results after 58 seconds of simulation for a thermo-mechanical problem with feedback from a micro-structure calculation that shows the temperature at the engineering scale, the geometry of each degraded micro-structure, and the effective thermal conductivity over time for each of the micro-structure calculations.}
    \label{fig:example}
\end{figure}

\gls{MOOSE} is capable of solving complex multiphysics problems, including systems with highly variable space and time scales. For example, the results in Figure~\ref{fig:example} couple an engineering scale problem to a micro-structure scale calculation. The engineering scale physics is porous flow, modeled using Darcy's equation within a cylinder assuming a porous media of closely packed steel spheres, mimicking the experiment detailed in \cite{pamuk2012friction}. Each end of the cylinder is attached to a vessel with a prescribed temperature and pressure. The simulation solves the mass, energy, and momentum balance equations for pressure, temperature, and thermally induced strain, respectively using a fully-coupled Newton-based solve. At each time step, a set of micro-structure calculations is performed, resolving the closely-packed spheres.  These spheres are degrading, causing an increase in permeability over time. The degradation rate increases with temperature, resulting in a change to the effective thermal conductivity at the engineering scale. Detailed steps for the solve are as follows:

\begin{enumerate}
\item At time $t=-1$ the solution for temperature is initialized to 300C, pressure, and displacements are set to zero; the effective thermal conductivity is set to the known value for close-packed spheres of steel in water.
\item The engineering scale problem is solved with a time step of 0.25 sec. (until $t=0$). During these steps, the temperature and pressure boundary conditions are increased linearly to 350C and 4000Pa, respectively.
\item\label{itm:one}The engineering problem computes a solution for a time step, samples the temperature at six evenly spaced points along the domain, and sets the temperature of the corresponding micro-structure calculations.
\item\label{itm:two}The micro-structure application computes degradation of the steel spheres using the temperature from the engineering scale application and then calculates the effective thermal conductivity based on the changed geometry. The resulting effective thermal conductivity is then applied using linear interpolation to the engineering-scale heat conduction calculation.
\item Steps \ref{itm:one}--\ref{itm:two} are repeated until a steady-state solution is reached.
\end{enumerate}

\section{Impact}

Software development is increasingly becoming an integral part of the scientific process. Currently, scientists and engineers develop their own software, spending approximately 30\% of their time writing code \cite{wilson2014best}. These domain scientists often do not follow code reuse best practices \cite{morisio2002success}, which can result in continuous reinvention of software as new researchers inherit opaque works left by a previous generation. By providing a framework that follows best practices for scientific computing, it is possible to aid researchers in combating these shortcomings to improve the quality of scientific research. Developing trust in a framework goes beyond utilizing best practices; it is equally important that the framework itself is built on a foundation of trusted tools. For this reason \gls{MOOSE} relies on the well-established code bases of libMesh~\cite{kirk2006libmesh} for finite elements and on PETSc~\cite{petsc-web-page, petsc-user-ref, petsc-efficient} for leading-edge numerical methods/solvers. Furthermore, \gls{MOOSE} itself has been subjected to multiple peer-reviews and \gls{sqa} audits.

\gls{MOOSE} is a powerful open-source, massively parallel, finite element based, multiphysics framework. Other open-source frameworks certainly exist (e.g., \cite{rathgeber2016firedrake, alnaes2015fenics, mfem-library}), but \gls{MOOSE} is a common platform for creation of comprehensive multiphysics applications that are inherently shareable and can be coupled without modification. \gls{MOOSE} is being used by governments, private industry, and universities both nationally and internationally. For example, \gls{MOOSE} is used for coupling multiple nuclear energy codes, both external and \gls{MOOSE}-based \cite{martineau2019multiphysics, ortensi2015full, dehart2016research, dehart2015multi}; it is used for modeling porous flow in 3D fractured porous media \cite{schadle20193d}; and is the basis of multiple efforts to model sub-surface exploration \cite{ennis2018community, pouletcsiro, regenauer2016next, lesueur2017framework}.

\section{Conclusions}
\label{sec:closing}

\gls{MOOSE} is a software framework that allows scientists to develop state-of-the-art, scalable applications without worrying about parallel, finite element, or solver implementation details. Moreover, the modular design of \gls{MOOSE} allows these applications to be combined easily, encouraging reuse and simplifying construction of multiscale, multiphysics simulations. The framework has attracted a sizable research community, pushing the frontiers of scientific computing, solving \gls{pde}'s with tens of billions of unknowns on tens of thousands of processors. \gls{MOOSE} development continues to focus on robust, high-performance solution algorithms that work across broad scales, from serial to hundreds of thousands of processes.

\section*{Acknowledgements}
\label{sec:acknowldegments}

This work was funded under multiple programs and organizations: The Idaho National Laboratory LDRD program, and The Department of Energy Nuclear Energy Advanced Modeling and Simulation and Consortium for Advanced Simulation of Light Water Reactors programs. This research made use of the resources of the High Performance Computing Center at the INL. This manuscript has been authored by Battelle Energy Alliance, LLC under Contract No.~DE-AC07-05ID14517 with the US Department of Energy.
The United States Government retains and the publisher, by accepting the article for publication, acknowledges that the United
States Government retains a nonexclusive, paid-up, irrevocable, worldwide license to publish or reproduce the published form
of this manuscript, or allow others to do so, for United States Government purposes.



\clearpage
\printglossary[type=\acronymtype]

%% References: At least 5 are required
\bibliographystyle{elsarticle-num}
\bibliography{moose}


%% \begin{thebibliography}{00}


%% \bibitem{}

%% \end{thebibliography}

\clearpage
\section*{Required Metadata}
\label{sec:metadata:required}

\section*{Current code version}
\label{sec:version:code}

%Ancillary data table required for subversion of the codebase. Kindly replace examples in right column with the correct information about your current code, and leave the left column as it is.

\begin{table}[!h]
\begin{tabular}{|l|p{6.5cm}|p{6.5cm}|}
\hline
\textbf{Nr.} & \textbf{Code metadata description} & \textbf{Please fill in this column} \\
\hline
C1 & Current code version & 0.9.0-pre \\
\hline
C2 & Permanent link to code/repository used of this code version &
\url{https://github.com/idaholab/moose} \\
\hline
C3 & Legal Code License & LGPL 2.1 \\
\hline
C4 & Code versioning system used & git \\
\hline
C5 & Software code languages, tools, and services used & C++, Python \\
\hline
C6 & Compilation requirements, operating environments \& dependencies &
\makecell[tl]{GCC 4.8.4+ Clang 3.5.1+ \\
Memory: 16GB+ \\
Disk: 30GB+ \\
OS: Mac OS 10.13+, Linux (POSIX) \\
Deps: MPI, PETSc, Hypre, libMesh} \\
\hline
C7 & If available Link to developer documentation/manual & \url{https://mooseframework.org} \\
\hline
C8 & Support email for questions & \url{moose-users@googlegroups.com} \\
\hline
\end{tabular}
\caption{Code metadata (mandatory)}
\label{sec:code:metadata}
\end{table}

\end{document}